\documentclass[10.8pt]{article}%
\usepackage{amsfonts}
\usepackage{amsmath}
\usepackage{amssymb}
\usepackage{charter}
\usepackage{graphicx}
\usepackage{cite}
\usepackage{geometry}
\providecommand{\U}[1]{\protect \rule{.1in}{.1in}}

\RequirePackage{CJK}
\AtBeginDocument{\begin{CJK*}{GBK}{song}\CJKtilde}
\AtEndDocument{\end{CJK*}}

\begin{document}

\title{Non-Abelian photon}

\author{Xiang-Yao Wu$^{a}$\thanks{E-mail: wuxy2066@163.com},
 Hong-Li$^{a}$, Xiao-Jing Liu$^{a, b}$, Si-Qi Zhang$^{a}$,
Guang-Huai Wang$^{a}$\\ Ji Ma$^{a}$, Heng-Mei Li$^{c}$, Hong-Chun
Yuan$^{c}$, Hai-Xin Gao$^{d}$ and Jing-Wu Li$^{e}$
\\{\small a) Institute of Physics, Jilin Normal University,
Siping 136000 China}\\
{\small b) Institute of Physics, Jilin University, Changchun
130012 China }\\{\small c) College of Optoelectronic Engineering,
Changzhou Institute of Technology, Changzhou 213002,
China}\\{\small d) Institute of Physics, Northeast Normal
University, Changchun 130024 China}\\
{\small e) Institute of Physics, Xuzhou Normal University, Xuzhou
221000 China}} \maketitle

\begin{abstract}
In this paper, we have proposed $SU(2)$ non-Abelian
electromagnetism gauge theory. In the theory, photon has
self-interaction and interaction, which can explain photon
entanglement phenomenon in quantum information. Otherwise, we find
there are three kinds photons $\gamma^{+}$, $\gamma^{-}$ and
$\gamma^{0}$, and they have electric charge $+e_{\gamma}$,
$-e_{\gamma}$ and $0$, respectively, which are accordance with
some experiment results.
\newline

\textbf{Keywords}: QED; SU(2) gauge theory

\textbf{PACS:} 11.15.-q, 12.20.-m

\end{abstract}

\section{Introduction}

The concept of photon as the quanta of the electromagnetic field
dates back to the beginning of this century. In order to explain
the spectrum of black-body radiation, Planck postulated the
process of emission and absorption of radiation by atoms occurs
discontinuously in quanta, i.e., the emission of black-body was
energy quantization with value of $\hbar \omega$ \cite{s1}, In
1905, Einstein had arrived at a more drastic interpretation. From
a statistical analysis of the Planck radiation law and from the
energetics of the photoelectric effect he concluded that it was
not merely the atomic mechanism of emission and absorption of
radiation which is quantized, but that electromagnetic radiation
itself consists of photons. The Compton effect confirmed this
interpretation.

The foundations of a systematic quantum theory of field were laid by Dirac in
1927. From the quantization of the electromagnetic field one is naturally led
to the quantization of any classical field, the quanta of the field being
particles with well-defined properties. We have successfully quantized the
free Dirac electron, we would like to discuss the question of coupling the
Dirac electron to a spin-one Maxwell field. The resulting theory had been
called quantum electrodynamics, namely QED. Over the past decades, the quantum
electrodynamics (QED) has attracted a considerable scientific attention
\cite{s3,s4}. As we have already stated, QED is an Abelian gauge theory, which
based on a $U(1)$ gauge symmetry. In 1954, Yang and Mills \cite{s5} extended
the gauge principle to non-Abelian gauge symmetry, which based not on the
simple one-dimensional group $U(1)$ of electrodynamics, but on a
three-dimensional group, the group $SU(2)$ of isotopic spin conservation, in
the hope that this would become a theory of the strong interactions. In
particular, because the gauge group was non-Abelian there was a
self-interaction of the gauge bosons, and the $U(1)$ Abelian gauge theory
there was not a self-interaction.

Entanglement \cite{s6} is a unique feature of quantum theory
having no analogue in classical physics. Spontaneous parametric
down-conversion (SPDC) has been used as a source of entangled
photon pairs for more than two decades \cite{s7} and provides an
efficient way to generate non-classical states of light for
fundamental tests of nature \cite{s8,s9}, for quantum information
processing \cite{s10,s11,s12} or for quantum metrology \cite{s13}.
Entanglement between two photons emitted by SPDC can occur in one
or several possible degrees of freedom of light \cite{s14}, namely
polarization, transverse momentum and energy. At present, the
two-photon, three-photon and multi-photon entanglement have been
observed in experiment \cite{s15,s16}. The photon entanglement is
from photon self-interaction and the interaction among photons. In
order to study the photon entanglement, we have extended the
Abelian QED to the non-Abelian QED, which can describe the photon
self-interaction and the interaction among photons.

In this paper, we have proposed $SU(2)$ non-Abelian
electromagnetism gauge theory. In the theory, photon has
self-interaction and interaction, which can explain photon
entanglement phenomenon in quantum information. Otherwise, we find
there are three kinds photons $\gamma^{+}$, $\gamma^{-}$ and
$\gamma^{0}$, and they have electric charge $+e_{\gamma}$,
$-e_{\gamma}$ and $0$, respectively, which are accordance with
some experiment results.

\section{QED with Abelian $U(1)$ gauge theory}

In quantum theory, QED is an Abelian gauge theory. It is instructive to show
that the theory ban be derived by the Dirac free electron theory to be gauge
invariant and renormalizable. Consider the Lagrangian for a free electron
field $\psi(x)$
\begin{equation}
L_{0}=\bar{\psi}(x)(i\gamma^{\mu}\partial_{\mu}-m)\psi(x),\label{1}%
\end{equation}
the Dirac fields $\psi(x)$ and $\bar{\psi}(x)$ under the $U(1)$ local gauge
transformations
\begin{align}
\psi(x) &  \rightarrow \psi^{\prime-i\alpha(x)}\psi(x)\nonumber \\
\bar{\psi}(x) &  \rightarrow \bar{\psi}^{\prime i\alpha(x)}\bar{\psi
}(x)\label{2}%
\end{align}
where $\alpha(x)$ is a real number. The derivative term will now have a rather
complicated transformation
\begin{equation}
\bar{\psi}(x)\partial_{\mu}\psi(x)\rightarrow \bar{\psi}^{\prime}%
(x)\partial_{\mu}\psi^{\prime}(x)=\bar{\psi}(x)\partial_{\mu}\psi
(x)-i\bar{\psi}(x)\partial_{\mu}\alpha(x)\psi(x),\label{3}%
\end{equation}
The second tern spoils the invariance. We need to form a gauge-covariant
derivative $D_{\mu}$, to replace $\partial_{\mu}$, and $D_{\mu}\psi(x)$ will
have the simple transformation
\begin{equation}
D_{\mu}\psi(x)\rightarrow \lbrack D_{\mu}\psi(x)]^{\prime-i\alpha(x)}D_{\mu
}\psi(x),\label{4}%
\end{equation}
so that the combination $\bar{\psi}(x)D_{\mu}\psi(x)$ is gauge invariant. In
other words, the action of the covariant derivative on the field will not
change the transformation property of the field. This can be realized if we
enlarge the theory with a new vector field $A_{\mu}(x)$, the gauge field, and
form the covariant derivative as
\begin{equation}
D_{\mu}\psi(x)=(\partial_{\mu}+ieA_{\mu})\psi(x),\label{5}%
\end{equation}
where $e$ is a free parameter which we eventually will identify with electric
charge. Then the transformation law for the covariant derivative (\ref{4})
will be satisfied if the gauge field $A_{\mu}(x)$ has the transformation
\begin{equation}
A_{\mu}(x)\rightarrow A_{\mu}^{\prime}(x)=A_{\mu}(x)+\frac{1}{e}\partial_{\mu
}\alpha(x),\label{6}%
\end{equation}
Form (\ref{1}) we now have
\begin{equation}
L=\bar{\psi}i\gamma^{\mu}(\partial_{\mu}+ieA_{\mu})-m\bar{\psi}\psi,\label{7}%
\end{equation}
defining gauge field tensor $F_{\mu \nu}$ as
\begin{equation}
F_{\mu \nu}=\partial_{\mu}A_{\nu}-\partial_{\nu}A_{\mu},\label{8}%
\end{equation}
under a transformation (\ref{6}), the field tensor $F_{\mu \nu}$ is invariant,
and we can structure the Lagrangian of $U(1)$ gauge field
\begin{equation}
L_{A}=-\frac{1}{4}F_{\mu \nu}F^{\mu \nu},\label{9}%
\end{equation}
under a transformations (\ref{2}) and (\ref{6}), the invariant total
Lagrangian of QED is
\begin{equation}
L=\bar{\psi}i\gamma^{\mu}(\partial_{\mu}+ieA_{\mu})\psi-m\bar{\psi}\psi
-\frac{1}{4}F_{\mu \nu}F^{\mu \nu}.\label{10}%
\end{equation}
The following features of (\ref{10}) should be noted\newline(1) The photon is
massless because a $A_{\mu}A^{\mu}$ term is not gauge invariant and not
included in (\ref{10}).\newline(2) The Lagrangian of (\ref{10}) does not have
a gauge field self-interaction.\newline

\section{QED with non-Abelian $SU(2)$ gauge theory}

In 1954, Yang and Mills extended the gauge principle to non-Abelian symmetry,
it is $SU(2)$ transformation group of isotopic spin. In order to study the
photon entanglement, we have extended the Abelian QED to the non-Abelian QED.
In the following, we shall study electromagnetism interaction with the $SU(2)$
gauge theory. We know electrons and positrons with the same mass and opposite
charges which obey the same equation. The Dirac equation must therefore admit
a new symmetry corresponding to the interchange particle and antiparticle. We
thus seek a transformation $\psi \rightarrow \psi^{c}$ reversing the charge, and
obtain the Dirac equations of electrons and positrons in electromagnetism
field
\begin{equation}
\lbrack \gamma^{\mu}(\partial_{\mu}-ieA_{\mu})-m]\psi(x)=0,\label{11}%
\end{equation}
and
\begin{equation}
\lbrack \gamma^{\mu}(\partial_{\mu}+ieA_{\mu})-m]{\psi}^{c}(x)=0,\label{12}%
\end{equation}
where
\begin{equation}
\bar{\psi}^{c}(x)=\eta_{c}C\bar{\psi}^{T}(x),\label{13}%
\end{equation}%
\begin{equation}
C=i\gamma_{2}\gamma_{0}=\left(
\begin{array}
[c]{cc}%
0 & -i\sigma_{2}\\
-i\sigma_{2} & 0
\end{array}
\right)  ,\label{14}%
\end{equation}
where $\eta_{c}$ an arbitrary unobservable phase, generally taken as being
equal to unity.\newline

Let the fermion fields $\psi(x)$ and $\psi^{c}(x)$ of electrons and positrons
be electric charge doublet of $SU(2)$
\begin{equation}
\psi=\left(
\begin{array}
[c]{c}%
\psi \\
\psi^{c}%
\end{array}
\right)  .\label{15}%
\end{equation}
For the scalar fields of electric charge, they are described by plural fields
$\varphi(x)$ and $\varphi^{\ast}(x)$, the charge doublet of $SU(2)$ is
\begin{equation}
\psi=\left(
\begin{array}
[c]{c}%
\varphi \\
\varphi^{\ast}%
\end{array}
\right)  .\label{16}%
\end{equation}
For equation (\ref{15}), under an $SU(2)$ transformation, we have
\begin{equation}
\psi(x)\rightarrow \psi^{^{\prime}}(x)=e^{-iT^{i}\cdot \theta^{i}}%
\psi(x),\label{17}%
\end{equation}
where $T^{i}=\frac{1}{2}\sigma^{i}$, $\sigma^{i}(i=1,2,3)$ are the usual Pauli
matrices, satisfying
\begin{equation}
\lbrack \frac{\sigma_{i}}{2},\frac{\sigma_{j}}{2}]=i\epsilon_{ijk}\frac
{\sigma_{k}}{2}\hspace{0.3in}i,j,k=1,2,3,\label{18}%
\end{equation}
and $\theta=(\theta_{1},\theta_{2},\theta_{3})$ are the $SU(2)$ transformation
parameters. The free Lagrangian for electrons field $\psi(x)$
\begin{equation}
L_{0}={\bar{\psi}(x)}(i\gamma^{\mu}\partial_{\mu}-m)\psi(x),\label{19}%
\end{equation}
is invariant under the global $SU(2)$ symmetry with ${\theta_{i}}$ being
space-time independent. However, under the local symmetry transformation
\begin{equation}
\psi(x)\rightarrow \psi^{\prime}(x)=u(\theta)\psi(x),\label{20}%
\end{equation}
with
\begin{equation}
u(\theta)=e^{-iT^{i}\cdot \theta^{i}(x)},\label{21}%
\end{equation}
the free Lagrangian $L_{0}$ is no longer invariant because the derivative term
transforms as
\begin{equation}
\bar{\psi}(x)\partial_{\mu}\psi(x)\rightarrow \bar{\psi}^{\prime}%
(x)\partial_{\mu}\psi^{\prime}(x)=\bar{\psi}(x)\partial_{\mu}\psi(x)+\bar
{\psi}(x)u^{-1}(\theta)[\partial_{\mu}u(\theta)]\psi(x),\label{22}%
\end{equation}
To construct a gauge-invariant Lagrangian we follow a procedure
similar to that of the Abelian case. First we introduce three
vector gauge fields $A_{\mu }^{i}$, $i=1,2,3$ (i.e., there are
three kinds of photons) for the $SU(2)$ gauge group to form the
gauge-covariant derivative through the minimal coupling
\begin{equation}
D_{\mu}(x)=\partial_{\mu}+A_{\mu}(x),\label{23}%
\end{equation}
where
\begin{equation}
A_{\mu}(x)=-igA_{\mu}^{i}(x)T^{i},\label{24}%
\end{equation}
where $g$ is the coupling constant in analogy to $e$ in (\ref{5}). We demand
that $D_{\mu}\psi$ have the same transformation property as $\psi$ itself,
i.e.
\begin{equation}
D_{\mu}\psi \rightarrow(D_{\mu}\psi)^{\prime}=D_{\mu}^{\prime}\psi^{\prime
}=D_{\mu}^{\prime}u(x)\psi=u(\theta)D_{\mu}\psi,\label{25}%
\end{equation}
This implies that
\begin{equation}
(\partial_{\mu}-igT^{i}A_{\mu}^{\prime i})(u(\theta)\psi)=u(\theta
)(\partial_{\mu}-igT^{i}A_{\mu}^{i})\psi,\label{26}%
\end{equation}
or
\begin{equation}
\lbrack \partial_{\mu}u(\theta)-igT^{i}A_{\mu}^{\prime i}u(\theta
)]=-igu(\theta)T^{i}A_{\mu}^{i},\label{27}%
\end{equation}
or
\begin{equation}
T^{i}A_{\mu}^{\prime i}=u(\theta)T^{i}A_{\mu}^{i}u^{-1}(\theta)-\frac{i}%
{g}[\partial_{\mu}u(\theta)]u^{-1}(\theta),\label{28}%
\end{equation}
which defines the transformation law for the gauge field. For an infinitesimal
gauge change $\theta(x)\ll1$,
\begin{equation}
u(\theta)\cong1-i\vec{T}\cdot \vec{\theta}(x),\label{29}%
\end{equation}
ignoring the higher order terms of $\theta^{j}$, equation (\ref{28}) becomes
\begin{align}
T^{i}A_{\mu}^{\prime i} &  =(1-iT^{j}\theta^{j})T^{i}A_{\mu}^{i}%
(1+iT^{j}\theta^{j})-\frac{i}{g}(-i\vec{T}\cdot \partial_{\mu}\vec{\theta
})(1+i\vec{T}\cdot \vec{\theta})\nonumber \\
&  =T^{i}A_{\mu}^{i}+iT^{i}A_{\mu}^{i}(T^{j}\theta^{j})-i(T^{j}\theta
^{j})T^{i}A_{\mu}^{i}-\frac{1}{g}(\vec{T}\cdot \partial_{\mu}\vec{\theta
}),\nonumber \\
&  =T^{i}A_{\mu}^{i}+iA_{\mu}^{k}\theta^{j}T^{k}T^{j}-iA_{\mu}^{k}\theta
^{j}T^{j}T^{k}-\frac{1}{g}(\vec{T}\cdot \partial_{\mu}\vec{\theta})\nonumber \\
&  =T^{i}A_{\mu}^{i}-iA_{\mu}^{k}\theta^{j}[T^{j},T^{k}]-\frac{1}{g}(\vec
{T}\cdot \partial_{\mu}\vec{\theta})\nonumber \\
&  =T^{i}A_{\mu}^{i}+A_{\mu}^{k}\theta^{j}\epsilon^{ijk}T^{i}-\frac{1}{g}%
(\vec{T}\cdot \partial_{\mu}\vec{\theta})\label{30}%
\end{align}
or
\begin{equation}
A_{\mu}^{\prime i}=A_{\mu}^{i}+\epsilon^{ijk}A_{\mu}^{k}\theta^{j}-\frac{1}%
{g}\partial_{\mu}\theta^{i},\label{31}%
\end{equation}
defining gauge field intensity $F_{\mu \nu}^{\alpha}$, it is
\begin{equation}
F_{\mu \nu}=D_{\mu}A_{\nu}-D_{\nu}A_{\mu}\label{32}%
\end{equation}
and
\begin{equation}
F_{\mu \nu}=-igF_{\mu \nu}^{\alpha}T^{\alpha},\label{33}%
\end{equation}
and
\begin{align}
-igF_{\mu \nu}^{\alpha}T^{\alpha} &  =(\partial_{\mu}-igA_{\mu}^{\alpha
}T^{\alpha})(-igA_{\nu}^{\beta}T^{\beta})-(\partial_{\nu}-igA_{\nu}^{\beta
}T^{\beta})(-igA_{\mu}^{\alpha}T^{\alpha})\nonumber \\
&  =-ig\partial_{\mu}A_{\nu}^{\beta}T^{\beta}+ig\partial_{\nu}A_{\mu}^{\alpha
}T^{\alpha}-g^{2}A_{\mu}^{\alpha}A_{\nu}^{\beta}[T^{\alpha},T^{\beta
}]\nonumber \\
&  =-ig(\partial_{\mu}A_{\nu}^{\alpha}-\partial_{\nu}A_{\mu}^{\alpha
})T^{\alpha}-ig^{2}\epsilon^{\alpha \beta \nu}A_{\mu}^{\alpha}A_{\nu}^{\beta
}T^{\nu},\label{34}%
\end{align}
or
\begin{equation}
F_{\mu \nu}^{\alpha}=\partial_{\mu}A_{\nu}^{\alpha}-\partial_{\nu}A_{\mu
}^{\alpha}+g\epsilon^{\alpha \beta \nu}A_{\mu}^{\beta}A_{\nu}^{\alpha
}.\label{35}%
\end{equation}
From equation (\ref{28}), we have
\begin{equation}
A_{\mu}^{\prime}(x)=u(\theta)A_{\mu}(x)u^{-1}(\theta)+u(\theta)\partial_{\mu
}u^{-1}(\theta),\label{36}%
\end{equation}
$F_{\mu \nu}^{\prime}$ gauge transformation is
\begin{align}
F_{\mu \nu}^{\prime} &  =D_{\mu}^{\prime}A_{\nu}^{\prime}-D_{\nu}^{\prime
}A_{\mu}^{\prime}\nonumber \\
&  =(\partial_{\mu}+A_{\mu}^{\prime})A_{\nu}^{\prime}-(\partial_{\nu}+A_{\nu
}^{\prime})A_{\mu}^{\prime}\nonumber \\
&  =(\partial_{\mu}+uA_{\mu}u^{-1}+u\partial_{\mu}u^{-1})(uA_{\nu}%
u^{-1}+u\partial_{\nu}u^{-1})\nonumber \\
&  -(\partial_{\nu}+uA_{\nu}u^{-1}+u\partial_{\nu}u^{-1})(uA_{\mu}%
u^{-1}+u\partial_{\mu}u^{-1}),\label{37}%
\end{align}
the first term is
\begin{align}
&  (\partial_{\mu}+uA_{\mu}u^{-1}+u\partial_{\mu}u^{-1})(uA_{\nu}%
u^{-1}+u\partial_{\nu}u^{-1})\nonumber \\
&  =(\partial_{\mu}u)A_{\nu}u^{-1}+u(\partial_{\mu}A_{\nu})u^{-1}+uA_{\nu
}(\partial_{\mu}u^{-1})+(\partial_{\mu}u)\partial_{\nu}u^{-1}+u\partial_{\mu
}\partial_{\nu}u^{-1}\nonumber \\
&  +uA_{\mu}u^{-1}uA_{\nu}u^{-1}+uA_{\mu}u^{-1}u\partial_{\nu}u^{-1}%
+u(\partial_{\mu}u^{-1})A_{\nu}u^{-1}+u(\partial_{\mu}u^{-1})u\partial_{\nu
}u^{-1}\nonumber \\
&  =u(\partial_{\mu}A_{\nu})u^{-1}+uA_{\nu}(\partial_{\mu}u^{-1}%
)+u\partial_{\mu}\partial_{\nu}u^{-1}+uA_{\mu}A_{\nu}u^{-1}+uA_{\mu}%
\partial_{\nu}u^{-1}\label{38}%
\end{align}
and the second term is
\begin{align}
&  (\partial_{\nu}+uA_{\nu}u^{-1}+u\partial_{\nu}u^{-1})(uA_{\mu}%
u^{-1}+u\partial_{\mu}u^{-1})\nonumber \\
&  =u(\partial_{\nu}A_{\mu})u^{-1}+uA_{\mu}(\partial_{\nu}u^{-1}%
)+u\partial_{\nu}\partial_{\mu}u^{-1}+uA_{\nu}A_{\mu}u^{-1}+uA_{\nu}%
\partial_{\mu}u^{-1}\label{39}%
\end{align}
substituting Eqs. (\ref{38}) and (\ref{39}) into (\ref{37}), we have
\begin{align}
F_{\mu \nu}^{\prime} &  =u[(\partial_{\mu}+A_{\mu})A_{\nu}-(\partial_{\nu
}+A_{\nu})A_{\mu}]u^{-1}\nonumber \\
&  =u(D_{\mu}A_{\nu}-D_{\nu}A_{\mu})u^{-1}\nonumber \\
&  =uF_{\mu \nu}u^{-1},\label{40}%
\end{align}
under an infinitesimal gauge change (29), there is
\begin{equation}
F_{\mu \nu}\rightarrow F_{\mu \nu}^{\prime}=(1-iT^{a}\theta^{a})F_{\mu \nu
}(1+iT^{b}\theta^{b}),\label{41}%
\end{equation}
or
\begin{align}
F_{\mu \nu}^{\prime c}T^{c} &  =(1-iT^{a}\theta^{a})F_{\mu \nu}^{c}%
T^{c}(1+iT^{b}\theta^{b})\nonumber \\
&  =F_{\mu \nu}^{c}T^{c}+iF_{\mu \nu}^{c}T^{c}T^{b}\theta^{b}-iF_{\mu \nu}%
^{c}T^{a}T^{c}\theta^{a}+F_{\mu \nu}^{c}T^{a}T^{c}\theta^{a}T^{b}\theta
^{b}\nonumber \\
&  =F_{\mu \nu}^{c}T^{c}+iF_{\mu \nu}^{c}[T^{c},T^{b}]\theta^{b}\nonumber \\
&  =F_{\mu \nu}^{c}T^{c}-\epsilon^{abc}F_{\mu \nu}^{a}T^{c}\theta^{b},\label{42}%
\end{align}
or
\begin{equation}
F_{\mu \nu}^{\prime c}=F_{\mu \nu}^{c}-\epsilon^{abc}F_{\mu \nu}^{a}\theta
^{b},\label{43}%
\end{equation}
i.e.,
\begin{equation}
F_{\mu \nu}^{\prime a}=F_{\mu \nu}^{a}-\epsilon^{cba}F_{\mu \nu}^{c}\theta
^{b}=F_{\mu \nu}^{a}+\epsilon^{abc}F_{\mu \nu}^{c}\theta^{b}.\label{44}%
\end{equation}
From equation (\ref{40}), we have
\begin{equation}
F_{\mu \nu}^{\prime}F^{\mu \nu^{\prime}}=uF_{\mu \nu}u^{-1}uF^{\mu \nu}%
u^{-1}=uF_{\mu \nu}F^{\mu \nu}u^{-1},\label{45}%
\end{equation}
and
\begin{equation}
TrF_{\mu \nu}^{\prime}F^{\mu \nu^{\prime}}=TruF_{\mu \nu}F^{\mu \nu}%
u^{-1}=TrF_{\mu \nu}F^{\mu \nu},\label{46}%
\end{equation}
or
\begin{align}
TrF_{\mu \nu}F^{\mu \nu^{\prime}} &  =Tr[(-igF_{\mu \nu}^{\alpha}T^{\alpha
})(-igF^{\mu \nu \beta}T^{\beta})]\nonumber \\
&  =-g^{2}F_{\mu \nu}^{\alpha}F^{\mu \nu \beta}Tr(T^{\alpha}T^{\beta})\nonumber \\
&  =-\frac{1}{2}g^{2}F_{\mu \nu}^{\alpha}F^{\mu \nu \beta}\delta_{\alpha \beta
}\nonumber \\
&  =-\frac{1}{2}g^{2}F_{\mu \nu}^{\alpha}F^{\mu \nu \alpha},\label{47}%
\end{align}
the Lagrangian of gauge field can be taken as
\begin{equation}
L_{F}=-\frac{1}{4}F_{\mu \nu}^{\alpha}F^{\mu \nu \alpha},\label{48}%
\end{equation}
and the total Lagrangian is
\begin{align}
L &  =\bar{\psi}(i\gamma^{\mu}D_{\mu}-m)\psi-\frac{1}{4}F_{\mu \nu}^{\alpha
}F^{\mu \nu \alpha}\nonumber \\
&  =\bar{\psi}(i\gamma^{\mu}\partial_{\mu}-m)\psi-g\bar{\psi}\gamma^{\mu
}A_{\mu}^{\alpha}T^{\alpha}\psi-\frac{1}{4}F_{\mu \nu}^{\alpha}F^{\mu \nu \alpha
}.\label{49}%
\end{align}
The $U(1)$ and $SU(2)$ gauge theory of Abelian and non-Abelian
have been introduced in many quantum field theory books
\cite{s17,s18,s19}. In this paper, we apply the non-Abelian gauge
theory to describe the electromagnetic interaction among photons,
and obtain some new results:\newline(1) The photon is massless
because a $A_{\mu}A^{\mu}$ term is not gauge invariant and not
included in (\ref{49}).\newline(2) The Lagrangian
of (\ref{49}) has a photon self-interaction, because of the term $\frac{1}%
{4}F_{\mu \nu}^{\alpha}F^{\mu \nu \alpha}$ in (\ref{49}) contains
products of three and four factors of $A_{\mu}$, and these imply
diagrams of the three-photon vertex and four-photon vertex, which
are shown in Fig. 1 (a) and (b).\newline(3) In $SU(2)$ gauge
theory, there are three kinds photons $\gamma^{+}$, $\gamma^{-}$
and $\gamma^{0}$, they have three different electric charge
$+e_{\gamma}$, $-e_{\gamma}$ and $0$, respectively. The electric
charge quantity of photon is more less than the electric's, i.e.,
$e_\gamma/e\ll 1$. In Refs. \cite{s20,s21,s22}, these experiments
have given the ratio of photon electric charge and electron
electric charge $e_\gamma/e < 3.4\times 10^{-5}$.

\begin{figure}[ptb]
\label{Fig1}
\centering \includegraphics[width=0.8\columnwidth]{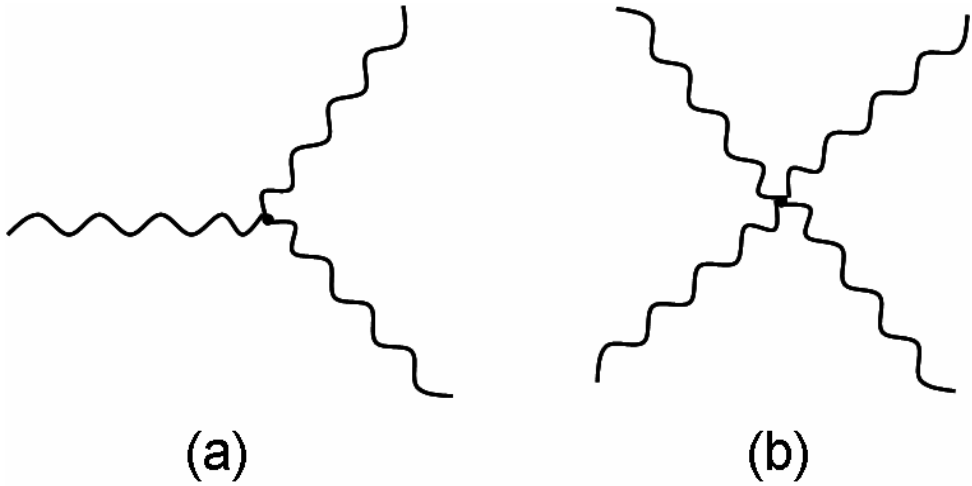}\caption{{\protect \small (a)
three-photon vertex \hspace{0.1in}(b) four-photon vertex}}%
\end{figure}

\section{Conclusion}

In this paper, we have proposed $SU(2)$ electromagnetism theory,
which is the gauge theory of two-dimension electric charge space.
In the theory, photon is massless, and it has self-interaction and
interaction, i.e., there are three-photon and four-photon vertex.
Otherwise, we find there are three kinds electric charge photons
$\gamma^{+}$, $\gamma^{-}$ and $\gamma^{0}$. In some experiments,
authors have found photon with teeny electric charge. We think the
theory should be further tested by experiment.


\begin{thebibliography}{99}                                                                                               %


\bibitem {s1}M. Planck, The Theory of Heat Radiation, Philadel- phia, (1914).

\bibitem {s2}A. Einstein, Concerning a Heuristic Point of View to-ward the
Emission and Transformation of Light, Annals of Physics, Vol. 17,
(1905).

\bibitem {s3}A. Di Piazza, C. Muller, K.Z. Hatsagortsyan, and C.H. Kietel,
Rev. Mod. Phys. 84, 1177 (2012).

\bibitem {s4}V.I. Ritus, J. Sov. Laser Res. 6, 497 (1985).

\bibitem {s5}C. N. Yang and R. L. Mills, Phys. Rev. 96, 191 (1954).

\bibitem {s6}R. Horodecki, P. Horodecki, M. Horodecki and K. Horodecki, Rev.
Mod. Phys. 81, 865 (2009).

\bibitem {s7}R. Ghosh and L. Mandel, Phys. Rev. Lett. 59, 1903 (1987).

\bibitem {s8}A. Zeilinger, Rev. Mod. Phys. 71, 288 ( 1999).

\bibitem {s9}M. Genovese, Phys. Rep. 413, 319 (2005).

\bibitem {s10}N. Gisin, G. Ribordy, W. Tittel and H. Zbinden, Rev. Mod. Phys.
74, 145 (2002).

\bibitem {s11}N. Gisin and R. Thew, Nature Photonics 1, 165 (2007).

\bibitem {s12}P. Kok, W. J. Munro, K. Nemoto, T. C. Ralph, J. P. Dowling and
G. J. Milburn, Rev. Mod. Phys. 79, 135 (2007).

\bibitem {s13}V. Giovannetti, S. Lloyd and L. Maccone, Nature Photonics 5, 222 (2011).

\bibitem {s14}J. T. Barreiro, N. K. Langford, N. A. Peters and P. G. Kwiat
Phys. Rev. Lett. 95, 260501 (2005).

\bibitem {s15}D. Bouwmeester, J-W Pan, Phys. Rev. Lett, 82, 1345 (1999).

\bibitem {s16}C. A. Sackett, Nature, 404, 256 (2000).

\bibitem {s17}J. B. Bjorken and S. D. Drell. Relativistic Quantum Fields. Mc-Graw-Hill, (1965).

\bibitem {s18}C. Itzykson and J. B. Zuber. Quantum Field Theory.
Mc-Graw-Hill, (1980).

\bibitem {s19}D. B. Lichtenberg, Unitary symmetry and elementary particles (2nd
edn). Academic Press, New York. (1978).

\bibitem {s20}M. I. Dobroliubov and A. Y. Ignatiev, Phys. Rev.
Lett. 65, 679 (1990).

\bibitem {s21}T. Mitsui, R. Fujimoto, Y. Ishisaki, Y. Ueda, Y.
Yamazaki, S. Asai and S. Orito, Phys. Rev. Lett. 70, 2265 (1993).

\bibitem {s22}A. Badertscher et al., Phys. Rev. D 75, 032004 (2007).

\end{thebibliography}
\end{document}